\documentclass[review,12pt]{elsarticle}

\usepackage{amssymb}
\usepackage{amsmath}
\usepackage{booktabs} 
\usepackage{multirow}
\usepackage{xcolor}
\newcommand{\differencecolor}{black}  
\newcommand{\differencecolorr}{black}  
\usepackage[ruled,vlined]{algorithm2e}
\usepackage{float}

\journal{Neural Networks}

\begin{document}

\begin{frontmatter}


\newcommand{\modelname}{NeuroDyGait}
\title{\textbf{EEG-to-Gait Decoding via Phase-Aware Representation Learning}}


\author[1]{Xi Fu}
\ead{fuxi0010@e.ntu.edu.sg}

\author[2]{Weibang Jiang}
\ead{935963004@sjtu.edu.cn}

\author[1]{Rui Liu}  
\ead{rui.liu@ntu.edu.sg}

\author[3]{Gernot R. M\"uller-Putz}  
\ead{gernot.mueller@tugraz.at}

\author[1,4]{Cuntai Guan \corref{cor1}}  
\ead{ctguan@ntu.edu.sg}
\address[1]{College of Computing and Data Science, Nanyang Technological University, Singapore 639798}
\address[2]{Department of Computer Science and Engineering, Shanghai Jiao Tong University,
Shanghai 200240, China}
\address[3]{Institute of Neural Engineering, Graz University of Technology, Graz, Austria}
\address[4]{Centre of AI in Medicine (C-AIM), Nanyang Technological University, Singapore}

\cortext[cor1]{Corresponding author}

\begin{abstract}
Accurate decoding of lower-limb motion from EEG signals is essential for advancing brain-computer interface (BCI) applications in movement intent recognition and control. 
\textcolor{\differencecolor}{This study presents \textbf{NeuroDyGait}, a two-stage, phase-aware EEG-to-gait decoding framework that explicitly models temporal continuity and domain relationships.}
To address challenges of causal, phase-consistent prediction and cross-subject variability, 
\textcolor{\differencecolor}{Stage I learns semantically aligned EEG–motion embeddings via relative contrastive learning with a cross-attention-based metric, while Stage II performs domain relation–aware decoding through dynamic fusion of session-specific heads.}
\textcolor{\differencecolor}{Comprehensive experiments on two benchmark datasets (GED and \textcolor{\differencecolor}{FMD}) show substantial gains over baselines, including a recent 2025 model EEG2GAIT.}
\textcolor{\differencecolor}{The framework generalizes to unseen subjects and maintains inference latency below 5 ms per window, satisfying real-time BCI requirements.}
\textcolor{\differencecolor}{Visualization of learned attention and phase-specific cortical saliency maps further reveals interpretable neural correlates of gait phases.}
\textcolor{\differencecolor}{Future extensions will target rehabilitation populations and multimodal integration.}
\end{abstract}


\end{frontmatter}


\newcommand{\modelname}{NeuroDyGait}
\section{Introduction}
Restoring human mobility using brain-computer interfaces (BCIs) is an emerging focus in neurorehabilitation~\cite{colucci2022brain}. A key component of this effort is the decoding of lower-limb motor intent from non-invasive electroencephalography (EEG) signals~\cite{he2018brain}. This capability is essential for enabling assistive technologies such as exoskeletons and neuroprosthetics~\cite{kilicarslan2021neuro}. Although recent progress has demonstrated the feasibility of EEG-based gait decoding, significant challenges remain for real-world deployment. A major limitation lies in the limited availability of normal motion data from patients with neurological impairments. These individuals are often unable to perform standard gait patterns. This makes it critical to design decoders that generalize across individuals without relying on patient-specific fine-tuning, thereby enabling practical and scalable use in rehabilitation settings~\cite{azab2019weighted, fahimi2019inter}.

Despite this progress, two key challenges hinder reliable decoding of gait dynamics. The first limitation concerns the widespread adoption of segment-to-point prediction strategies. These methods use a short window of historical EEG signals to estimate the motion state at a single final time step~\cite{he2018mobile, brantley2018full, shafiul2020prediction, presacco2011neural}.This formulation satisfies the requirements of causal decoding. However, it reduces human motion to isolated time points, neglecting the temporal continuity and biomechanical constraints that characterize natural gait. In practice, the context of gait provides critical information for inferring joint angles, as joint configurations are strongly influenced by surrounding temporal dynamics. Human locomotion is intrinsically structured and phase-dependent: even under similar motor intent, joint trajectories can vary significantly across individuals, and identical joint configurations may reflect different gait phases depending on context and prior states~\cite{hamacher2017exploring, presacco2012decoding}. Without explicitly modeling this structure, predictions risk becoming unstable or semantically inconsistent. Therefore, a robust decoding framework must incorporate these underlying dynamics to enable the model to differentiate between gait states that are superficially similar but functionally distinct across different cycles.

The second challenge lies in the limitations of current cross-domain learning strategies, which often treat multiple sessions or subjects as independent sources. These methods typically use shared encoders for generalization or subject-specific heads for individual differences~\cite{jayaram2016transfer, chen2021ms}, but fail to capture structured relationships across domains. This oversimplification becomes particularly problematic when generalizing across heterogeneous data sources, where domain shifts—caused by factors such as sensor placement, individual differences, and recording conditions—introduce structured dependencies between domains. Empirical evidence indicates that both inter-session and inter-subject variability can substantially degrade the consistency and reliability of biosignal decoding. This highlights the importance of models that not only capture domain-specific characteristics but also exploit shared structure across domains to enable more robust and transferable representation learning~\cite{huang2023discrepancy, maswanganyi2022statistical, ng2023deep}. Therefore, a robust framework should be designed to model both intra-domain specificity and inter-domain structure to enable more adaptive and generalizable learning.

To address these challenges, we propose \modelname, a domain-generalizable EEG-to-gait decoding framework built on semantically structured representation learning and relational domain modeling. The name reflects the model’s focus on neural (Neuro) dynamics (Dy) underlying continuous human locomotion (Gait), emphasizing its ability to capture temporally evolving brain-motor relationships for robust cross-subject prediction. Our contributions are twofold:

\begin{itemize}

\item \textbf{Phase-Aware Pretraining with Relative Contrastive Learning:}
We propose a novel pretraining strategy based on relative contrastive learning that aligns EEG and motor embeddings by modeling sample-wise semantic similarity. Unlike traditional contrastive approaches that rely on predefined positive-negative pairs, our method employs a learnable distance function to infer relative similarity among all samples within a batch. This relative similarity modeling approach enables the model to learn fine-grained, phase-aware gait semantics and distinguish between motion states that are functionally distinct but visually similar. Furthermore, the model is trained to reconstruct multi-cycle gait trajectories from EEG embeddings, which helps it capture neuromusculoskeletal dynamics in context and enhances its robustness to noisy or atypical inputs.

\item \textbf{Relation-Aware Decoding via Domain Head Mixture:}
We propose a domain-aware decoding mechanism that assigns a dedicated prediction head to each source session. During fine-tuning, a learnable domain head dynamically integrates intra-domain outputs with a weighted combination of cross-domain predictions, enabling the model to capture structured inter-session dependencies. This adaptive fusion of domain-specific knowledge enhances generalization across subjects by leveraging latent relationships among heterogeneous signal patterns, thereby enabling more robust and context-aware decoding in real-world scenarios.
\end{itemize}
\textcolor{\differencecolor}{
In summary, NeuroDyGait adopts a two-stage EEG-to-gait decoding framework: Stage~I learns phase-aware, semantically structured EEG–motion embeddings through multi-objective pretraining, and Stage~II performs relation-aware decoding by dynamically fusing session-specific heads based on learned inter-domain attention. This design enables the model to capture both temporal gait dynamics and structured cross-domain relationships for robust cross-subject prediction. We have released the full implementation as open-source at https://github.com/FuXi1999/NeuroDyGait.}

\section{Related Work}

\subsection{EEG-Based Motor Execution Decoding}

EEG-based neural decoding of lower-limb motor execution has been widely studied, leveraging EEG's high temporal resolution for capturing motor control dynamics. Pfurtscheller and colleagues~\cite{pfurtscheller2001motor} demonstrated that event-related desynchronization (ERD) and synchronization (ERS) effectively characterize motor execution and imagery, laying the groundwork for decoding lower-limb movements. Ang and colleagues~\cite{ang2012filter} used common spatial pattern (CSP) and linear discriminant analysis (LDA) to decode gait phases, enhancing motor-related EEG feature extraction. Schirrmeister et al.~\cite{schirrmeister2017deep} employed convolutional neural networks (CNNs), boosting accuracy in classifying lower-limb motor imagery. Recent deep learning methods have further improved spatiotemporal feature extraction from EEG signals~\cite{goh2018spatio, tortora2020deep, fu2022matn}, demonstrating their ability to capture complex nonstationary neural patterns. Finally, Wang et al.~\cite{wang2018implementation} developed a real-time BCI system for lower-limb exoskeleton control, validating the feasibility of EEG-driven motor decoding in rehabilitation settings. While these works significantly advance EEG-based BCIs, they typically focus on classifying discrete movement states or rely on pointwise regression, often lacking biomechanical awareness and dynamic consistency in continuous motion decoding.

\subsection{Cross-Subject Generalization in EEG Decoding}

A major challenge in EEG decoding lies in inter-subject variability, where differences in brain dynamics, head geometry, and sensor placement cause distributional shifts between individuals. To address this, domain adaptation methods have been applied to align source and target feature distributions through adversarial training~\cite{li2018cross}, statistical moment matching~\cite{long2015learning}, or subspace projection~\cite{liang2022multi}. However, most approaches assume access to target-domain data during training, which is often unrealistic in clinical settings where normative data from impaired users are unavailable. Domain generalization methods aim to overcome this by learning domain-invariant representations solely from source data~\cite{wang2024dmmr, lu2023hybrid}, yet often treat all source domains independently or equally. In practice, EEG and motor signals vary both across and within subjects, and ignoring inter-session relationships can limit generalization.

In this work, we model session-level structure explicitly by assigning each source session a dedicated decoder and learning a dynamic mixture of their outputs. This relation-aware strategy captures both intra-session specificity and cross-session similarity, improving robustness in unseen-subject transfer.

\subsection{Contrastive Representation Learning for Structured Movement}
Contrastive learning has proven effective for self-supervised representation learning in structured movement tasks like gait analysis and motor imagery decoding. By comparing similarities and differences between samples, it captures discriminative features without needing labels.

In video-based action recognition, motion-aware frameworks such as MaCLR align visual and motion modalities for enhanced video representations~\cite{xiao2022maclr}, while MCL emphasizes motion through alignment of gradient maps and optical flow~\cite{li2021motion}. For skeleton-based action recognition, contrastive learning captures structural dynamics~\cite{guo2022contrastive}, with HiCLR enforcing consistency across hierarchical augmentations~\cite{zhang2023hierarchical}, and cross-modality approaches modeling complex patterns~\cite{li2023cross}.

In EEG-based motor imagery, contrastive learning addresses inter-subject variability using CNNs and attention mechanisms~\cite{li2024self}, while supervised contrastive learning improves gait recognition using EEG and EMG signals~\cite{fu2023gait}.

These developments highlight the strength of contrastive learning in modeling dynamic, structured movement across various modalities.

\section{Methods}
\textcolor{\differencecolor}{To improve clarity and reproducibility, Table \ref{tab:notation} summarizes all symbols and parameters used in the framework.}

\textcolor{\differencecolor}{
\begin{table}[htbp]
\centering
\resizebox{0.95\linewidth}{!}{
\begin{tabular}{ll}
\toprule
\textbf{Symbol} & \textbf{Definition} \\
\midrule
$\mathbf{x} \in \mathbb{R}^{C \times T}$ & EEG signal segment (C channels, T time points) \\
$\mathbf{y} \in \mathbb{R}^{J \times T}$ & Motion (joint angle) sequence (J joints, T frames) \\
$f_e(\cdot)$, $f_m(\cdot)$ & EEG encoder and motor encoder networks \\
$g(\cdot)$ & Motor decoder network \\
$\mathbf{z}_e$, $\mathbf{z}_m$ & EEG and motor latent embeddings \\
$\hat{\mathbf{y}}$, $\hat{\mathbf{y}}_{T}$ & Reconstructed motion sequence and final predicted frame \\
$N$ & Number of samples per mini-batch \\
$d$ & Latent embedding dimension \\
$W_q, W_k, W_v, W_o \in \mathbb{R}^{d \times d}$ & Query, key, value, and output projection matrices in cross-attention \\
$\eta$ & Cross-attention coefficient (Stage I) \\
$\tau$ & Temperature parameter in contrastive learning \\
$b$ & Learnable bias for similarity scaling \\
$d(\cdot, \cdot)$ & Cross-modal distance measure between EEG and motor embeddings \\
$\mathcal{L}_{\text{rec}}$, $\mathcal{L}_{\text{pred}}$, $\mathcal{L}_{\text{rcl}}$ & Reconstruction, prediction, and relative contrastive losses \\
$h_s(\cdot)$ & Session-specific prediction head for domain/session $s$ \\
$\mathcal{L}_{\text{sup}}$, $\mathcal{L}_{\text{df}}$ & Supervised and domain-fusion losses in Stage II \\
$\boldsymbol{\alpha} \in [0,1]^{N_{\text{src}}}$ & Domain weighting vector predicted by scoring network (Stage II) \\
$W_a$, $\mathbf{b}_a$ & Parameters of the domain scoring network \\
$\mathbf{m} \in \{0,1\}^{N_{\text{src}}}$ & One-hot mask excluding current session \\
$\hat{\mathbf{y}}_{\text{mix}}$, $\hat{\mathbf{y}}_{\text{test}}$ & Mixture predictions during training and inference \\
\bottomrule
\end{tabular}
}
\caption{Summary of main notations used in the proposed framework.}
\label{tab:notation}
\end{table}
}

\subsection{Overview}

Our proposed framework, \modelname, consists of two training stages: 
(1) a pretraining stage (\textcolor{\differencecolor}{Stage I}) that learns temporally structured and semantically meaningful EEG embeddings by reconstructing synchronized motion signals and modeling cross-domain relationships, optimized using a combination of reconstruction, contrastive, and prediction losses; and
(2) a domain generalization stage (\textcolor{\differencecolor}{Stage II}) that employs a session-wise head architecture with a learnable domain fusion mechanism, optimized via a domain fusion loss and a supervised prediction loss.

\begin{figure*}[htbp]
    \centering
    \includegraphics[width=0.95\linewidth]{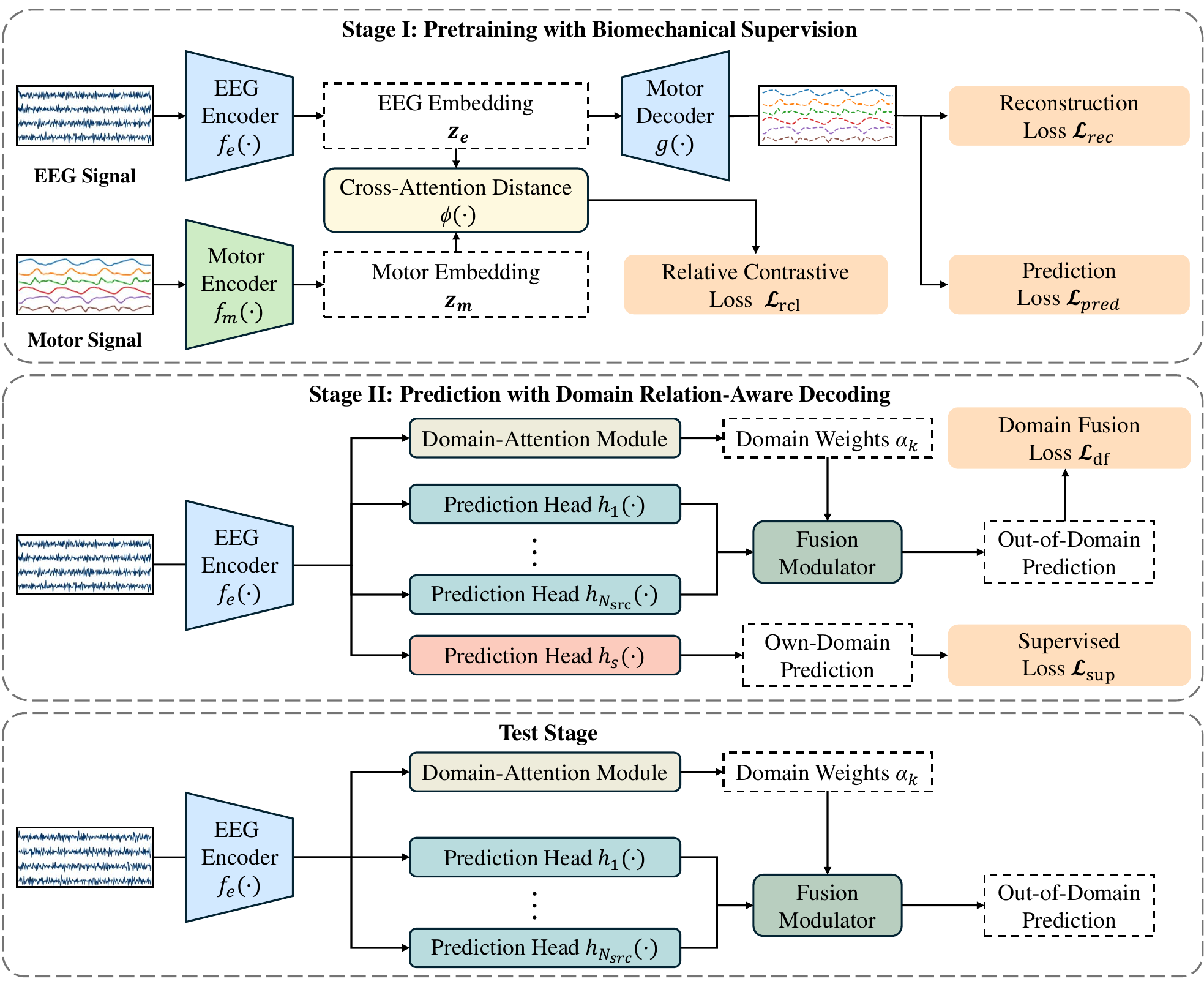}
    \caption{Overview of \modelname\ architecture. \textbf{Stage I}: The dual-encoder model consists of an EEG encoder, a motor encoder, and a decoder. It is trained with reconstruction, contrastive, and prediction losses to extract biomechanically meaningful EEG representations. \textbf{Stage II}: During training, session-specific heads and a domain weighting layer are optimized to predict final motor state. \textbf{Test}: The model computes a normalized mixture of all source-domain heads for unseen-domain generalization.}
    \label{model}
\end{figure*}

An overview of \modelname, including the Stage I, Stage II, and Test Stage, is shown in Fig.~\ref{model}.

\subsection{Stage I: Pretraining with Biomechanical Supervision}

\subsubsection{Dual-Encoder Architecture}

The pretraining stage employs two modality-specific encoders: an EEG encoder $f_e$ and a motor encoder $f_m$. EEG signals within 2-second windows are processed by $f_e$ to produce embeddings $\mathbf{z}_e = f_e(\mathbf{x})$, while synchronized joint angle sequences are passed through $f_m$ to produce $\mathbf{z}_m = f_m(\mathbf{y})$. A motor decoder $g$ reconstructs the motion signal $\hat{\mathbf{y}} = g(\mathbf{z}_e)$ from the EEG embeddings.

The EEG encoder is composed of a deep convolutional feature extractor that captures multi-scale spatiotemporal patterns across channels, followed by a final constrained $1 \times T$ convolution layer to project the output into a compact latent space. The use of filter growth (25, 50, 100, 200) across layers promotes hierarchical abstraction, and norm-constrained weights ensure stability in training.

The motor encoder combines temporal convolutional blocks with a Transformer module. Specifically, stacked 1D convolutions reduce temporal resolution while enriching local features, and a Transformer encoder equipped with positional encoding captures long-range temporal dependencies in the joint trajectories. The resulting sequence representation is aggregated via average pooling to obtain a fixed-length latent embedding.

The motor decoder $g$ reconstructs time-domain joint angle sequences $\hat{\mathbf{y}} = g(\mathbf{z}_e)$ from EEG embeddings using a transposed CNN architecture. It first projects the latent vector into a spatiotemporal tensor, then applies a series of transposed convolutional layers with batch normalization and ReLU activations. The final layer outputs 6-dimensional joint angles, and dynamic output padding ensures the sequence length matches the original (400 time steps). This structure supports end-to-end training and preserves temporal continuity in motion reconstruction.

\subsubsection{Reconstruction Loss}

To ensure the EEG embedding captures biomechanically meaningful information, we train the decoder to reconstruct the full 2-second motion sequence from EEG embeddings using mean squared error (MSE):

\begin{equation}
\mathcal{L}_{\text{rec}} = \frac{1}{N} \sum_{i=1}^{N} \|\hat{\mathbf{y}}_i - \mathbf{y}_i\|^2
\end{equation}
where $N$ denotes the number of training samples in a batch.

This loss encourages the EEG representation to retain sufficient information for accurately capturing motor dynamics over time.

\subsubsection{Prediction Loss}

While the reconstruction loss encourages the EEG embeddings to preserve the full temporal dynamics of motion, our ultimate objective is to accurately predict the gait state at the final time point of the sequence. To this end, we introduce a prediction loss that focuses explicitly on the final frame of the motor output:

\begin{equation}
\mathcal{L}_{\text{pred}} = \frac{1}{N} \sum_{i=1}^{N} \|\hat{\mathbf{y}}_{i, T} - \mathbf{y}_{i, T}\|^2
\end{equation}
where $\hat{\mathbf{y}}_{i, T}$ and $\mathbf{y}_{i, T}$ denote the predicted and ground truth motor states at the final time step $T$ for the $i$-th sample.

This loss encourages the EEG embedding to preserve information that is most relevant for predicting the motor state at the last frame, thereby reducing the potential influence of earlier, less relevant gait phases during pretraining.

\subsubsection{Relative Contrastive Learning}

Inspired by Xu et al.~\cite{xu2024relcon}, we introduce a relative contrastive learning framework that captures fine-grained semantic relationships between EEG and motor signals. Unlike traditional contrastive methods that define fixed positive and negative sets, our approach treats all intra-batch samples as candidates and enforces relative similarity rankings between them. This encourages the model to learn an embedding space where the distance structure reflects the semantic alignment between modalities.

\paragraph{Cross-Attention-Based Distance}
Given an EEG embedding $\mathbf{z}_e \in \mathbb{R}^d$ and a candidate motor embedding $\mathbf{z}_m \in \mathbb{R}^d$, we define a cross-attention mechanism that treats $\mathbf{z}_e$ as the query and $\mathbf{z}_m$ as the key and value. The attention coefficient is computed via scaled dot-product:

\begin{equation}
\textcolor{\differencecolor}{\eta = \text{softmax}\!\left( \frac{\langle W_q \mathbf{z}_e, W_k \mathbf{z}_m \rangle}{\sqrt{d}} \right)}
\end{equation}

The attended motor feature is computed as:
\begin{equation}
\hat{\mathbf{z}}_\text{m} = W_o (\textcolor{\differencecolor}{\eta} \cdot W_v \mathbf{z}_m)
\end{equation}
The cross-modal distance is then defined as:
\begin{equation}
d(\mathbf{z}_e, \mathbf{z}_m) = \|\hat{\mathbf{z}}_m - \mathbf{z}_m\|_2^2
\end{equation}

\paragraph{Relative Contrastive Loss}
Using pairwise distances $d(\mathbf{z}_e^{(i)}, \mathbf{z}_m^{(j)})$ between all EEG–motor pairs in a batch, we define a relative contrastive loss enforcing ranking consistency:
\begin{equation}
S_{ij} = -\frac{d(\mathbf{z}_e^{(i)}, \mathbf{z}_m^{(j)})}{\tau} + b
\end{equation}
For each anchor, we compute the listwise loss:
\begin{equation}
\mathcal{L}_i = -\sum_{j=1}^{N-1} \left( S_{ij} - \log \sum_{k=j}^{N-1} e^{S_{ik}} \right)
\end{equation}
and the total contrastive loss:
\begin{equation}
\mathcal{L}_{\text{rcl}} = \frac{1}{N} \sum_{i=1}^{N} \mathcal{L}_i
\end{equation}

The final Stage I objective is:
\textcolor{\differencecolor}{
\begin{equation}
\mathcal{L}_{\text{stage I}} = \mathcal{L}_{\text{rec}} +  \mathcal{L}_{\text{pred}} + \mathcal{L}_{\text{rcl}}
\end{equation}
}
\paragraph{Loss Interaction Rationale}
\textcolor{\differencecolor}{All Stage I losses are equally weighted, as this configuration achieves stable convergence and balanced gradients without explicit tuning. Combining sequence reconstruction and final-frame prediction encourages the encoder to capture temporal continuity beneficial for Stage II fine-tuning rather than overfitting to instantaneous frames.}

\subsection{Stage II: Prediction with Domain Relation-Aware Decoding}

\subsubsection{Session-Specific Prediction Heads}

In Stage II, we retain only the pretrained EEG encoder and attach a prediction head $h_s$ to each source-domain session. For a given EEG embedding $\mathbf{z}_e$ belonging to session $s$, the corresponding session-specific head $h_s$ outputs a prediction $\hat{\mathbf{y}} = h_s(\mathbf{z}_e)$ for the final motor frame. Each session head is trained using the MSE loss:
\begin{equation}
\mathcal{L}_{\text{sup}} = \frac{1}{N} \sum_{i=1}^{N} \|\hat{\mathbf{y}}_i - \mathbf{y}_i\|^2
\end{equation}

\subsubsection{Domain Weighting Layer and Mixture Prediction}

To leverage inter-session relationships during inference, we introduce a domain weighting mechanism that adaptively fuses predictions from all source-domain heads except the one corresponding to the input session~\cite{yaoimproving}. Given an EEG embedding \( \mathbf{z}_e \in \mathbb{R}^d \) from session \( s \), the mixture prediction is computed as:

\begin{equation}
\hat{\mathbf{y}}_{\text{mix}} = \sum_{k=1}^{N_{\text{src}}} \alpha_k(\mathbf{z}_e) \cdot h_k(\mathbf{z}_e)
\end{equation}

where \( h_k(\cdot) \) denotes the domain-specific head for session \( k \), and \( \alpha_k(\mathbf{z}_e) \in [0, 1] \) is the attention weight predicted by a lightweight scoring network:

\begin{equation}
\boldsymbol{\alpha} = \text{softmax} \left( W_a \mathbf{z}_e + \mathbf{b}_a + \log(1 - \mathbf{m}) \right)
\end{equation}

Here, \( W_a \in \mathbb{R}^{N_{\text{src}} \times d} \) and \( \mathbf{b}_a \in \mathbb{R}^{N_{\text{src}}} \) are learnable parameters, and \( \mathbf{m} \in \{0,1\}^{N_{\text{src}}} \) is a one-hot mask where \( m_s = 1 \) denotes the current session. The additive masking term \( \log(1 - \mathbf{m}) \) sets the attention logit of the input session to \( -\infty \), ensuring that \( \alpha_s = 0 \). This excludes the intra-domain head from the mixture, promoting the learning of a cross-domain feature aggregation function.

\subsubsection{Joint Optimization with $\mathcal{L}_{\text{sup}}$ and $\mathcal{L}_{\text{df}}$}

The session heads and domain weighting mechanism are jointly optimized with:
\begin{equation}
\mathcal{L}_{\text{df}} = \frac{1}{N} \sum_{i=1}^N \|\hat{\mathbf{y}}_{\text{mix},i} - \mathbf{y}_i\|^2
\end{equation}
and the total Stage II loss:
\textcolor{\differencecolor}{
\begin{equation}
\mathcal{L}_{\text{stage II}} = \mathcal{L}_{\text{sup}} + \mathcal{L}_{\text{df}}
\end{equation}
}
\subsubsection{Inference via Head Mixture}

During inference on an unseen domain, we discard all session identity information and compute the final prediction as a weighted combination over all available source-domain heads:

\begin{equation}
\hat{\mathbf{y}}_{\text{test}} = \sum_{i=1}^{N_{\text{src}}} \alpha_i \cdot h_i(\mathbf{z}_e^{tgt})
\end{equation}
where weights $\alpha_i$ are predicted by the domain weighting layer. This strategy enables unseen-domain generalization by leveraging learned inter-session relationships without requiring any target-domain adaptation.

\textcolor{\differencecolorr}{For clarity, Algorithm~\ref{alg:framework} provides the pseudocode of the complete two-stage training and inference process.}

\begin{algorithm}[H]
\caption{Two-Stage Training and Inference Pipeline of \modelname}
\label{alg:framework}
\SetAlgoLined
\DontPrintSemicolon

\KwIn{EEG windows $X$, joint-angle labels $Y$, source domains $\{\mathcal{D}_1,\ldots,\mathcal{D}_{N_{\text{src}}}\}$}
\KwOut{Predicted joint-angle trajectory $\hat{Y}$}

\textbf{Stage I: Phase-Aware Pretraining}\;

\ForEach{batch $(X, Y)$}{
    $Z \leftarrow f_{\theta}(X)$\tcp*{EEG encoder}
    $\hat{X} \leftarrow g_{\phi}(Z)$\tcp*{Reconstruction}
    $\hat{H} \leftarrow p_{\psi}(Z)$\tcp*{Kinematic prediction}

    Compute relative similarity scores $s(Z_i,Z_j)$\;
    Compute contrastive loss $\mathcal{L}_{\text{rcl}}$\;

    Compute total Stage~I loss
    $\mathcal{L}_{\text{I}} = \|X - \hat{X}\|^2 + \|H - \hat{H}\|^2 + \mathcal{L}_{\text{rcl}}$\;

    Update $(\theta,\phi,\psi)$ via gradient descent\;
}

\BlankLine
\textbf{Stage II: Domain Relation-Aware Decoding}\;

Freeze encoder $f_{\theta}$\;

\ForEach{source domain $\mathcal{D}_k$}{
    Train domain-specific decoder $d_k$\;
}

Train attention module $A(\cdot)$ to output weights
$\alpha = A(Z)$\;

Final prediction:
$\hat{Y} = \sum_{k=1}^{N_{\text{src}}} \alpha_k\, d_k(Z)$\;

\BlankLine
\textbf{Inference}\;

\ForEach{incoming EEG window $X$}{
    $Z \leftarrow f_{\theta}(X)$\;
    $\alpha \leftarrow A(Z)$\;
    $\hat{Y} \leftarrow \sum_k \alpha_k\, d_k(Z)$\;
}

\Return{$\hat{Y}$}

\end{algorithm}
\section{Experiment}
\subsection{Dataset}
\subsubsection{Gait-EEG Dataset}

To investigate the brain mechanisms involved in walking, we collected a new dataset, Gait-EEG Dataset (GED)~\cite{fu2025eeg2gait}, recording brain activity along with simultaneous lower-limb joint angles natural walk on level ground. The dataset contains the recordings from 50 able-bodied participants (25 males, 25 females; aged 21 to 46, mean age 28.4, standard deviation 5.2), with no history of neurological disorders or lower limb pathologies. Participants engaged in two independent level-ground walking experiment sessions, with every session comprising three identical walking blocks. Each block included approximately 40 trials, with each trial representing EEG signals and synchronized lower-limb joint angles as the participant walked straight for 7.7 meters. Sessions were spaced at least three days apart. The dataset includes synchronized recordings from a 60-channel active EEG, a 4-channel electrooculogram (EOG), along with measurements from six joint angle sensors (bilateral hips, knees, and ankles)~\cite{fu2025eeg2gait}.

This study has been reviewed and approved by the Institutional Review Board (IRB-2021-709) of Nanyang Technological University, ensuring compliance with applicable legislation, ethical and safety requirements in Singapore. All participants have provided informed consents before the experiment. 

\subsubsection{Open-access Dataset}
To further validate our proposed method, we conducted additional experiments using the open-access Full body Mobile brain-body imaging Dataset (\textcolor{\differencecolor}{FMD})~\cite{brantley2018full}. The \textcolor{\differencecolor}{FMD} contains full-body motion capture data (66 markers) from approximately 10 walking trials performed by 10 able-bodied individuals under various gait conditions, including level ground, ramp, and stair walking. Data were collected using a wireless IMU-based motion capture system, enabling natural, unconstrained movement. To characterize lower-limb motor states, we selected the Z-axis joint angles of eight bilateral joints: hip, knee, ankle, and ball of the foot. A data segmentation stride of 50 ms was used.

\subsection{Data Preprocessing}

EEG signals from both datasets were preprocessed using an identical pipeline. A bandpass filter ranging from 0.1 to 48 Hz was first applied to eliminate low-frequency drifts and high-frequency noise. The filtered signals were then re-referenced using common average referencing (CAR) to reduce spatially correlated noise across channels. Subsequently, the EEG signals were resampled to 200 Hz to reduce computational overhead while preserving relevant neural activity. Similarly, goniometer signals were resampled to 200 Hz and normalized to zero mean and unit variance on a per-joint basis. Channels with zero standard deviation were safely handled by substituting a value of one to avoid division errors.

\subsection{Data Segregation}\label{segregation}

\subsubsection{GED}
\textcolor{\differencecolor}{For the Gait-EEG dataset, we adopted a 10-fold cross-subject evaluation to rigorously assess generalization. The 50 participants were evenly partitioned into ten disjoint folds, each containing data from five distinct subjects (e.g., subjects 1–5, 6–10, …, 46–50). In each round, one fold was used for testing, one for validation, and the remaining eight for training. This rotating scheme ensures that every subject serves as a test participant exactly once, providing a comprehensive measure of cross-subject robustness.}

\subsubsection{FMD}
\textcolor{\differencecolor}{For the \textcolor{\differencecolor}{FMD}, we followed a leave-one-subject-out (LOSO) protocol due to its smaller sample size of ten participants. In each iteration, data from nine subjects were used for training and one held-out subject for testing, while one additional subject was designated for validation within the training pool. Participant~1 was excluded from all folds because of EEG–motion desynchronization issues. This LOSO design enables an unbiased estimation of model generalization to completely unseen individuals.}

\textcolor{\differencecolor}{These subject-wise cross-validation settings emphasize the model’s capacity to generalize across individuals rather than trials, aligning with the goal of robust and transferable gait decoding.}

\subsubsection{Segment Statistics}
\textcolor{\differencecolor}{
To provide a clearer understanding of the data volume used for model training and evaluation, we quantified the number of segmented EEG–gait windows generated per session after preprocessing and sliding-window segmentation.}

\textcolor{\differencecolor}{For the GED dataset, each session contains on average \textbf{14,019.34} segments, with a standard deviation of \textbf{3,205.92} across all 100 sessions.  
For the \textcolor{\differencecolor}{FMD} dataset, each session contains on average \textbf{32,877.44} segments, with a standard deviation of \textbf{5,047.82} across all 9 valid subjects.}

\textcolor{\differencecolor}{These statistics provide a detailed view of the sample sizes available for the cross-subject evaluation protocols described above.
}
\subsection{Evaluation Metric}

We evaluated the efficacy of \modelname\ by comparing the predicted angles of joints with their actual recorded angles, using three standard regression metrics: Pearson correlation coefficient ($r$), coefficient of determination ($R^2$), and root mean squared error (RMSE). These metrics jointly capture trend consistency, explained variance, and absolute prediction accuracy.

\begin{equation}
r = \frac{\operatorname{cov}(y, \hat{y})}{\sigma(y) \cdot \sigma(\hat{y})}
\end{equation}

\begin{equation}
R^2 = 1 - \frac{\sum_{i=1}^n (y_i - \hat{y}_i)^2}{\sum_{i=1}^n (y_i - \bar{y})^2}
\end{equation}

\begin{equation}
\text{RMSE} = \sqrt{\frac{1}{n} \sum_{i=1}^{n} (y_i - \hat{y}_i)^2}
\end{equation}

Here, $y$ denotes the actual joint angle and $\hat{y}$ represents the predicted angle. The covariance between two variables $A$ and $B$ is denoted as $\operatorname{cov}(A, B)$, and $\sigma(A)$ is the standard deviation of $A$. $\bar{y}$ indicates the mean of the actual values. Each sequence represents data collected from a single trial over $n$ time steps, corresponding to $\frac{n}{\textcolor{\differencecolor}{200}}$ seconds at a sampling rate of \textcolor{\differencecolor}{200} Hz.

The Pearson $r$ value reflects the consistency in trend between the predicted and actual trajectories. The $R^2$ score measures the proportion of variance explained by the model, while RMSE captures the average magnitude of prediction error, penalizing larger deviations more heavily.

\subsection{Implementation and Hyperparameter Settings}

\modelname\ was implemented using the PyTorch library. Training was conducted using the Adam optimizer with default hyperparameter settings. A batch size of 512 was used, and training continued for a maximum of 50 epochs ($epoch_{\text{max}}$). Instead of using a fixed learning rate, we adopted a cosine learning rate scheduler with linear warm-up. The learning rate starts from zero and increases linearly during the initial warm-up phase, reaching the predefined maximum learning rate at the end of the warm-up period. After that, it gradually decreases to a minimum value following a cosine decay curve over the remaining training epochs. During the initial warm-up phase, spanning the first 2 epochs, the learning rate increases linearly from zero to the initial value. The total schedule is computed based on the number of training epochs and the number of iterations per epoch. The \textcolor{\differencecolor}{highest} learning rate and minimum learning rate are listed in Table~\ref{hyperparameter}.

\textcolor{\differencecolor}{For clarity and reproducibility, all dataset-specific quantities referenced in the equations are explicitly summarized here. For both GED and FMD, we use a sequence length of $T = 400$ samples (corresponding to 2 seconds at 200 Hz), $C = 64$ EEG channels, and $J = 6$ lower-limb joints. The number of source-domain sessions $N_{\text{src}}$ follows the leave-one-subject-out protocol: for GED, $N_{\text{src}} = 49$; for FMD, $N_{\text{src}} = 9$. All symbols and variable definitions are consolidated in Table~1 (Notation Table).}

\textcolor{\differencecolor}{Stage~I uses equal weighting for the reconstruction, prediction, and relative contrastive losses. As discussed in Section~6, jointly optimizing these objectives yields stable convergence without conflicting gradients. Sensitivity analysis further shows that varying the loss weights within moderate ranges results in less than a 2\% change in decoding performance, indicating robustness to the choice of loss weights. Likewise, varying the embedding dimension $d$ between 64 and 256 also produced less than a 2\% performance change, demonstrating low sensitivity to this hyperparameter.}

\begin{table}[t]
\centering
\caption{Training hyperparameters used in the cosine learning rate schedule for each stage and dataset.}
\label{hyperparameter}
\begin{tabular}{lcc}
\toprule
\textbf{Setting} & \textbf{Highest LR} & \textbf{Min LR} \\
\midrule
GED (Stage I)   & 1e-3  & 1e-4   \\
GED (Stage II)  & 2e-5  & 2e-6   \\
\textcolor{\differencecolor}{FMD (Stage I)}   & 1e-4  & 1e-5   \\
\textcolor{\differencecolor}{FMD (Stage II)}  & 5e-5  & 5e-6   \\
\bottomrule
\end{tabular}
\end{table}
\section{Results and Analysis}

In this section, we conduct a comprehensive evaluation of the proposed framework across multiple key dimensions. We begin by assessing model performance, comparing our method against several state-of-the-art baselines on two benchmark datasets: GED~\cite{fu2025eeg2gait} and \textcolor{\differencecolor}{FMD}~\cite{brantley2018full}. The results demonstrate consistent improvements in motion prediction accuracy, highlighting the effectiveness of our approach. We then examine cross-dataset transferability, evaluating the model’s generalization capability across different datasets, which underscores its robustness and adaptability in out-of-distribution settings. To further probe the model's internal behavior, we visualize phase-specific EEG embeddings using t-SNE, revealing structured, phase-aligned representations that reflect the temporal specificity of the learned features. Additionally, we analyze the relationship between domain attention entropy and out-of-domain predictive error, illustrating how the domain-aware decoding mechanism leverages inter-session dynamics to enhance decoding performance. We also perform spatial analysis via saliency mapping to identify cortical regions critical to the model’s predictions. The results indicate that the model predominantly focuses on central sensorimotor channels, such as Cz and CP2, consistent with the neural correlates of lower-limb motor control. Finally, we present a series of ablation studies to systematically assess the contribution of key architectural and training components to the overall performance of the proposed framework.

\subsection{Model Performance}

In this section, we validate the performance of \modelname\ on both the GED and \textcolor{\differencecolor}{FMD} datasets and compare it against a comprehensive set of state-of-the-art deep learning and machine learning approaches for EEG-based decoding. The baseline models include ContraWR~\cite{yang2021self}, FFCL~\cite{li2022motor}, TSception~\cite{ding2022tsception}, Temporal Convolutional Network (TCN)~\cite{ingolfsson2020eeg}, ST-Transformer~\cite{song2021transformer}, EEGConformer~\cite{song2023eeg}, SPaRCNet~\cite{jing2023development}, EEGNet~\cite{lawhern2018eegnet}, deepConvNet~\cite{schirrmeister2017deep}, and the recent state-of-the-art EEG2GAIT model~\cite{fu2025eeg2gait}.

\textcolor{\differencecolor}{We emphasize that EEG-to-gait regression is a highly specialized task, and only a limited number of domain-generalization techniques exist specifically for this setting. To ensure both fairness and relevance, we selected the strongest and most widely adopted EEG-based regression and representation-learning models currently available. In particular, the inclusion of EEGConformer (2023), SPaRCNet (2023), and EEG2GAIT (2025) ensures that our evaluation covers the most recent methodological advances in EEG decoding. These models collectively represent the current landscape of high-performing architectures for neural-based gait prediction, providing a rigorous benchmark for assessing the effectiveness of \modelname.}

The results of all evaluation metrics are reported to characterize \modelname's performance relative to existing methods on both datasets. \textcolor{\differencecolor}{\modelname\ achieved an $r$ value of 0.6980 ($R^2$ = 0.4847, RMSE = 0.7329) on GED and an $r$ value of 0.2945 ($R^2$ = 0.0485, RMSE = 0.9743) on FMD, consistently outperforming all baseline approaches.} All baseline models were trained following the strategy described in Section~\ref{segregation}, ensuring a fair comparison.

A summary of quantitative performance across methods is provided in Table~\ref{tab:GEDperformance} and Table~\ref{tab:FMDperformance}. Across both GED and \textcolor{\differencecolor}{FMD}, \modelname\ achieves the best or second-best performance on all metrics, demonstrating robust generalization ability and substantial improvements over established EEG-based decoding architectures.

\begin{table*}[htbp]
\centering
\caption{Performance comparison of different methods on the GED dataset (mean $\pm$ std across folds).}
\begin{tabular}{lccc}
\toprule
\textbf{Method} & \textbf{Pearson $r$} $\uparrow$ & \textbf{$R^2$ Score} $\uparrow$ & \textbf{RMSE} $\downarrow$ \\
\midrule
ContraWR~~\cite{yang2021self}              & 0.3227 $\pm$ 0.1263 & 0.0830 $\pm$ 0.0834 & 0.9857 $\pm$ 0.0477 \\
FFCL~~\cite{li2022motor}                   & 0.4793 $\pm$ 0.1020 & 0.1946 $\pm$ 0.1133 & 0.9217 $\pm$ 0.0663 \\
TSception~~\cite{ding2022tsception}        & 0.3664 $\pm$ 0.0793 & 0.1197 $\pm$ 0.0846 & 0.9653 $\pm$ 0.0471 \\
EEGNet~~\cite{lawhern2018eegnet}           & 0.4301 $\pm$ 0.1697 & 0.1917 $\pm$ 0.1476 & 0.9219 $\pm$ 0.0801 \\
TCN~~\cite{ingolfsson2020eeg}              & 0.2608 $\pm$ 0.0522 & 0.0621 $\pm$ 0.0373 & 0.9977 $\pm$ 0.0205 \\
EEGConformer~~\cite{song2023eeg}           & 0.5292 $\pm$ 0.0817 & 0.2698 $\pm$ 0.1005 & 0.8774 $\pm$ 0.0623 \\
SPaRCNet~~\cite{jing2023development}       & 0.6422 $\pm$ 0.0655 & 0.3756 $\pm$ 0.1037 & 0.8093 $\pm$ 0.0704 \\
ST-Transformer~~\cite{song2021transformer} & 0.5963 $\pm$ 0.0795 & 0.3471 $\pm$ 0.1097 & 0.8282 $\pm$ 0.0710 \\
deepConvNet~~\cite{schirrmeister2017deep}  & 0.6904 $\pm$ 0.0679 & 0.4773 $\pm$ 0.0955 & \underline{0.7392 $\pm$ 0.0717} \\
EEG2GAIT~~\cite{fu2025eeg2gait}            & \underline{0.6962 $\pm$ 0.0764} & \underline{0.4819 $\pm$ 0.1091} & 0.7387 $\pm$ 0.0969 \\
\textbf{\modelname}                        & \textbf{0.6980 $\pm$ 0.0742} & \textbf{0.4847 $\pm$ 0.1085} & \textbf{0.7329 $\pm$ 0.0809} \\
\bottomrule
\end{tabular}
\label{tab:GEDperformance}

\vspace{1mm}
\begin{minipage}{0.95\linewidth}
\footnotesize
\textit{↓: lower is better; ↑: higher is better. The best results are in bold; second best are underlined.}
\end{minipage}
\end{table*}
\begin{table*}[htbp]
\centering
\caption{Performance comparison of different methods on the \textcolor{\differencecolor}{FMD} (mean $\pm$ std across subjects).}
\begin{tabular}{lccc}
\toprule
\textbf{Method} & \textbf{Pearson $r$} $\uparrow$ & \textbf{$R^2$ Score} $\uparrow$ & \textbf{RMSE} $\downarrow$ \\
\midrule
ContraWR~~\cite{yang2021self}              & 0.1838 $\pm$ 0.0490 & 0.0291 $\pm$ 0.0278 & 0.9846 $\pm$ 0.0152 \\
EEGConformer~~\cite{song2023eeg}           & 0.2443 $\pm$ 0.0842 & -0.0514 $\pm$ 0.1132 & 1.0215 $\pm$ 0.0540 \\
EEGNet~~\cite{lawhern2018eegnet}           & 0.1827 $\pm$ 0.0986 & \underline{0.0256 $\pm$ 0.0280} & \underline{0.9861 $\pm$ 0.0153} \\
FFCL~~\cite{li2022motor}                   & 0.1545 $\pm$ 0.1122 & 0.0078 $\pm$ 0.0479 & 0.9947 $\pm$ 0.0233 \\
EEG2GAIT~~\cite{fu2025eeg2gait}            & 0.1984 $\pm$ 0.0966 & 0.0198 $\pm$ 0.0322 & 0.9886 $\pm$ 0.0174 \\
SPaRCNet~~\cite{jing2023development}       & 0.1811 $\pm$ 0.0349 & -0.0379 $\pm$ 0.0288 & 1.0175 $\pm$ 0.0145 \\
ST-Transformer~~\cite{song2021transformer} & 0.1716 $\pm$ 0.0258 & -0.0367 $\pm$ 0.0281 & 1.0172 $\pm$ 0.0134 \\
TCN~~\cite{ingolfsson2020eeg}              & 0.0709 $\pm$ 0.0312 & -0.0136 $\pm$ 0.0151 & 1.0061 $\pm$ 0.0086 \\
TSception~~\cite{ding2022tsception}        & 0.1698 $\pm$ 0.0512 & -0.3285 $\pm$ 0.6594 & 1.1229 $\pm$ 0.2229 \\
deepConvNet~~\cite{schirrmeister2017deep}  & \underline{0.2185 $\pm$ 0.0881} & -0.0313 $\pm$ 0.1034 & 1.0124 $\pm$ 0.0505 \\
\textbf{\modelname}                        & \textbf{0.2945 $\pm$ 0.1131} & \textbf{0.0485 $\pm$ 0.0342} & \textbf{0.9743 $\pm$ 0.0185} \\
\bottomrule
\end{tabular}
\label{tab:FMDperformance}

\vspace{1mm}
\begin{minipage}{0.95\linewidth}
\footnotesize
\textit{↓: lower is better; ↑: higher is better. The best results are in bold; second best are underlined.}
\end{minipage}
\end{table*}


\subsection{Cross-Dataset Transferability}
To evaluate the transferability of learned EEG representations, we conducted a cross-dataset experiment in which Stage~I of \modelname\ was first pretrained on GED. The resulting EEG encoder was then paired with a randomly initialized motor encoder and decoder adapted to the \textcolor{\differencecolor}{FMD} joint structure. This model served as the initialization for training on the \textcolor{\differencecolor}{FMD}, and was further trained through Stage~I and Stage~II using \textcolor{\differencecolor}{FMD} data only.

This approach tests whether the EEG encoder, trained on one dataset, can provide a better initialization than random weights for a new domain, thereby facilitating the model in learning motor-relevant features and improving performance. Table~\ref{tab:cross_dataset} presents the results of this transfer setup compared to training \modelname\ from scratch on \textcolor{\differencecolor}{FMD}. The transferred model achieves marginal improvements over training from scratch across all three metrics, suggesting better generalization and robustness of the EEG representations learned through cross-dataset pretraining.

\begin{table}[htbp]
\centering
\caption{Cross-dataset transfer results (GED $\rightarrow$ \textcolor{\differencecolor}{FMD}, mean $\pm$ std across subjects).}
\label{tab:cross_dataset}
\small  
\setlength{\tabcolsep}{6pt}  
\begin{tabular}{lccc}
\toprule
\textbf{Init. Strategy} & \textbf{Pearson $r$} $\uparrow$ & \textbf{$R^2$ Score} $\uparrow$ & \textbf{RMSE} $\downarrow$ \\
\midrule
Scratch (\textcolor{\differencecolor}{FMD} only)       & {0.2945 $\pm$ 0.1131} & {0.0485 $\pm$ 0.0342} & {0.9743 $\pm$ 0.0185}  \\
GED$\rightarrow$\textcolor{\differencecolor}{FMD} Init &{0.2983 $\pm$ 0.1036} & {0.0546 $\pm$ 0.0316} & {0.9717 $\pm$ 0.0147}  \\
\bottomrule
\end{tabular}

\vspace{1mm}
\begin{minipage}{0.9\linewidth}
\footnotesize
\textit{Initializing the EEG encoder from a Stage I model pretrained on GED improves performance on \textcolor{\differencecolor}{FMD}, demonstrating cross-dataset transferability.}
\end{minipage}
\end{table}

\subsection{Enhancing Performance via Target Domain Fine-Tuning}

Although \modelname\ outperforms or matches all baselines on both GED and \textcolor{\differencecolor}{FMD}, its predictive performance on \textcolor{\differencecolor}{FMD} remains relatively modest (Pearson $r = \textcolor{\differencecolor}{0.2945}$, $R^2 = \textcolor{\differencecolor}{0.0485}$), likely due to the small number of subjects and increased inter-subject variability.

To simulate realistic BCI deployment with limited calibration data, we performed target domain fine-tuning after training stage I and II by using only the first 3 minutes of EEG data per test session (less than 8\% of total session length). The first 2.5 minutes were used for fine-tuning, and the remaining 30 seconds for early stopping validation.

As summarized in Table~\ref{tab:finetune}, fine-tuning substantially improved performance on both datasets, with particularly pronounced gains on \textcolor{\differencecolor}{FMD}. On \textcolor{\differencecolor}{FMD}, Pearson $r$ increased from \textcolor{\differencecolor}{0.2945} to \textcolor{\differencecolor}{0.4202}, corresponding to an improvement of \textcolor{\differencecolor}{+0.1257}. In addition, $R^2$ rose from \textcolor{\differencecolor}{0.0485} to \textcolor{\differencecolor}{0.1812}, representing an increase of \textcolor{\differencecolor}{+0.1327}. Meanwhile, RMSE decreased by \textcolor{\differencecolor}{0.0771}. On GED, fine-tuning also enhanced all metrics, increasing $r$ from \textcolor{\differencecolor}{0.6980} to \textcolor{\differencecolor}{0.7906} (an improvement of \textcolor{\differencecolor}{+0.0926}) and reducing RMSE from \textcolor{\differencecolor}{0.7329} to \textcolor{\differencecolor}{0.6061} (a reduction of \textcolor{\differencecolor}{0.1268}).

\begin{table}[htbp]
\centering
\caption{Effect of fine-tuning using only the first 3 minutes of test-session EEG data (<8\% of the total session duration). Performance is summarized as mean~$\pm$~std across folds for GED and across subjects for \textcolor{\differencecolor}{FMD}.}

\begin{tabular}{lcccc}
\toprule
\textbf{Dataset} & \textbf{Metric} & \textbf{Original} & \textbf{Fine-Tuned} & \textbf{Improvement} \\
\midrule
\multirow{3}{*}{GED}
& Pearson $r$~($\uparrow$)  & 0.6980 $\pm$ 0.0742 & \textbf{0.7906 $\pm$ 0.1201} & +0.0926 \\
& $R^2$ Score~($\uparrow$)  & 0.4847 $\pm$ 0.1085 & \textbf{0.6304 $\pm$ 0.1733} & +0.1457 \\
& RMSE~($\downarrow$)       & 0.7329 $\pm$ 0.0809 & \textbf{0.6061 $\pm$ 0.1381} & -0.1268 \\
\midrule
\multirow{3}{*}{\textcolor{\differencecolor}{FMD}}
& Pearson $r$~($\uparrow$)  & 0.2945 $\pm$ 0.1131 & \textbf{0.4202 $\pm$ 0.1856} & +0.1257 \\
& $R^2$ Score~($\uparrow$)  & 0.0485 $\pm$ 0.0342 & \textbf{0.1812 $\pm$ 0.1748} & +0.1327 \\
& RMSE~($\downarrow$)       & 0.9743 $\pm$ 0.0185 & \textbf{0.8972 $\pm$ 0.1048} & -0.0771 \\
\bottomrule
\end{tabular}
\label{tab:finetune}
\end{table}

\subsection{t-SNE Visualization of Phase-Specific EEG Embeddings}



\subsubsection{Kinematic Basis for Four‐Phase Segmentation}\label{sec:kinematic_segmentation}
To inject gait‐cycle structure into our EEG embeddings, we partitioned each continuous gait cycle into four phases by detecting four reproducible kinematic events in the sagittal‐plane hip and knee trajectories of each leg~\cite{neptune2001contributions,anderson2004contributions,zajac2002biomechanics,whittle2014gait}:

\begin{enumerate}
\item[\textbf{(a)}] \emph{Left hip maximal flexion} (just prior to left toe‐off): marks end of left support and onset of left swing.
\item[\textbf{(b)}] \emph{Left knee maximal flexion} (mid‐swing peak): corresponds to peak elevation of the left limb for foot clearance.
\item[\textbf{(c)}] \emph{Right hip maximal flexion} (just prior to right toe‐off): marks end of right support and onset of right swing.
\item[\textbf{(d)}] \emph{Right knee maximal flexion} (mid‐swing peak): corresponds to peak elevation of the right limb.
\end{enumerate}

By chaining these events in temporal order \textbf{(a) → (b) → (c) → (d)}, we define four gait cycle phases:
\begin{itemize}
\item \textbf{Phase 1:} From (a) to (b)
\item \textbf{Phase 2:} From (b) to (c)
\item \textbf{Phase 3:} From (c) to (d)
\item \textbf{Phase 4:} From (d) to the next (a)
\end{itemize}

Each phase reflects a distinct interlimb coordination pattern and captures transitions between swing and stance for both legs.

\subsubsection{Visualization Method}
\begin{figure}[htbp]
    \centering
    \includegraphics[width=0.8\linewidth]{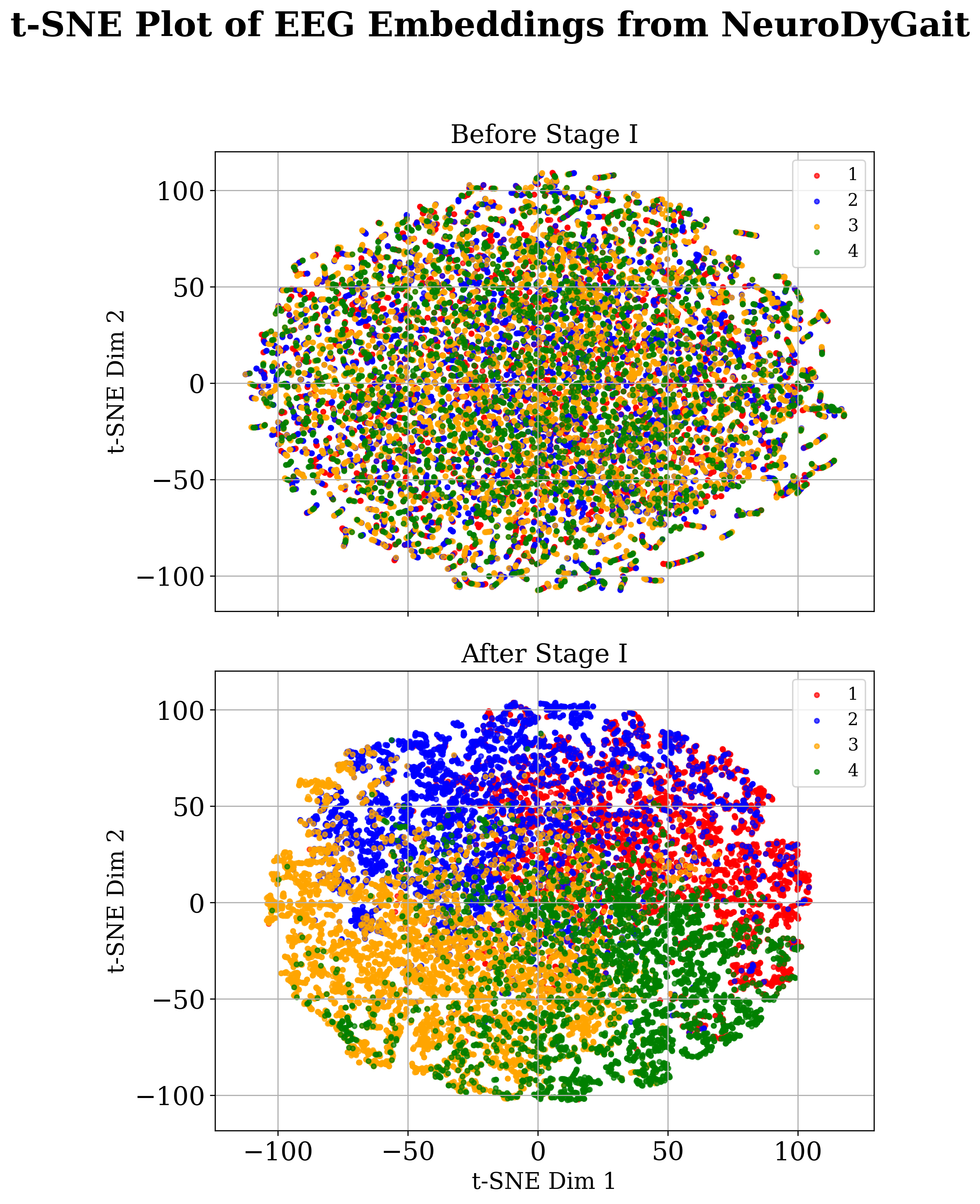}
    \caption{t-SNE visualization of EEG embeddings from before (upper panel) and after (lower panel) Stage~I training. Colors represent different gait phases as defined in Section\ref{sec:kinematic_segmentation}.}
    \label{tsne}
\end{figure}

Using the EEG encoder pretrained in Stage I, we extracted embeddings for overlapping EEG windows. Each window was labeled according to the phase of its final time‐sample kinematic phase and assigned a distinct color. We then projected all embeddings into two dimensions with t‐SNE~\cite{van2008visualizing}, and overlaid the phase‐color labels using Matplotlib’s \texttt{scatter()} (alpha=0.7). 
As shown in Fig.~\ref{tsne}, embeddings from different gait phases form spatially distinct clusters, demonstrating that the Phase‐Aware Pretraining objective implicitly captures the periodic structure of gait.

\subsection{Domain Attention Entropy and Predictive Performance}

To better understand the behavior of our domain fusion mechanism, we analyzed the entropy of the attention weights produced by the domain attention module for each session in the test set. Specifically, we computed the entropy of the predicted domain weight distribution at inference time, where a higher entropy indicates broader reliance on multiple source domains, while a lower entropy reflects more selective attention to a few sources.

\textcolor{\differencecolor}{In the updated analysis, entropy and decoding error were computed over all batches of test data across all cross-validation folds, providing a fold-aggregated view of model behavior.}

\begin{figure}[htbp]
    \centering
    \includegraphics[width=0.6\linewidth]{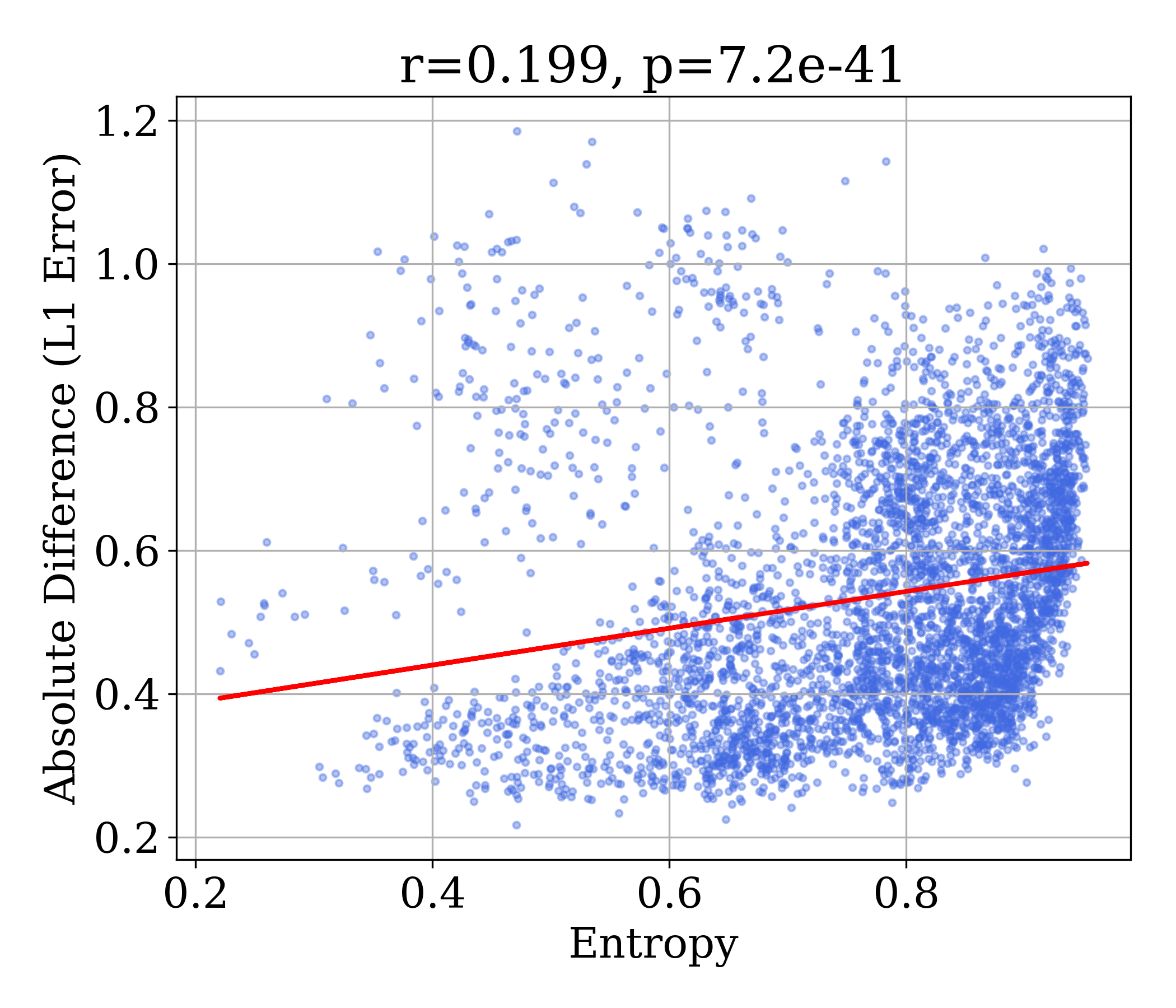}
    \caption{Scatter plot showing the relationship between domain attention entropy and L1 prediction error across test sessions. \textcolor{\differencecolor}{All results are aggregated over all cross-validation folds.}}
    \label{fig:entropy_vs}
\end{figure}

As shown in Fig.~\ref{fig:entropy_vs}, \textcolor{\differencecolor}{the aggregated results reveal a clear positive association between domain attention entropy and decoding error.} Each data point represents the mean entropy and L1 Error computed over a batch of test data. \textcolor{\differencecolor}{Across all folds, the Pearson correlation coefficient (PCC) is 0.199 with a $p \ll 1 \times 10^{-5}$, indicating a strong and statistically significant relationship.} \textcolor{\differencecolor}{A fitted regression curve computed over the aggregated dataset further illustrates that higher entropy is consistently linked to increased prediction error.}

\subsection{Spatial Analysis}
In addition to the performance metrics presented, we expanded our evaluation to include spatial feature importance analysis to better understand critical areas during decoding. To achieve this, we employed saliency mapping—a technique in machine learning that visualizes the importance of each input feature for the model's predictions~\cite{simonyan2013deep}. This method highlights the input areas the model is most sensitive to when making predictions. The saliency map, $S$, is generated by calculating the gradient of the model’s output with respect to each input feature. The gradients are visualized to represent how variations in each input element, $X_{ij}$ (where $i, j$ are the spatial and temporal indices of $X$ , an input sample in $\mathbf{R}^{C\times T}$), influence the output prediction. The magnitude of each element $S_{ij}$ in $S$ illustrates the importance of the corresponding input pixel $X_{ij}$ to the output prediction.

To derive a spatial saliency map from these calculations, we first averaged $S$ across the temporal dimension to obtain \( \Bar{S} \). We then projected \( \Bar{S} \) onto the corresponding scalp electrode positions, creating a topographical map that illustrates the focal areas of brain activity relevant to the model's decisions.

\begin{figure}[htbp]
  
    \centering
    \includegraphics[width=0.8\linewidth]{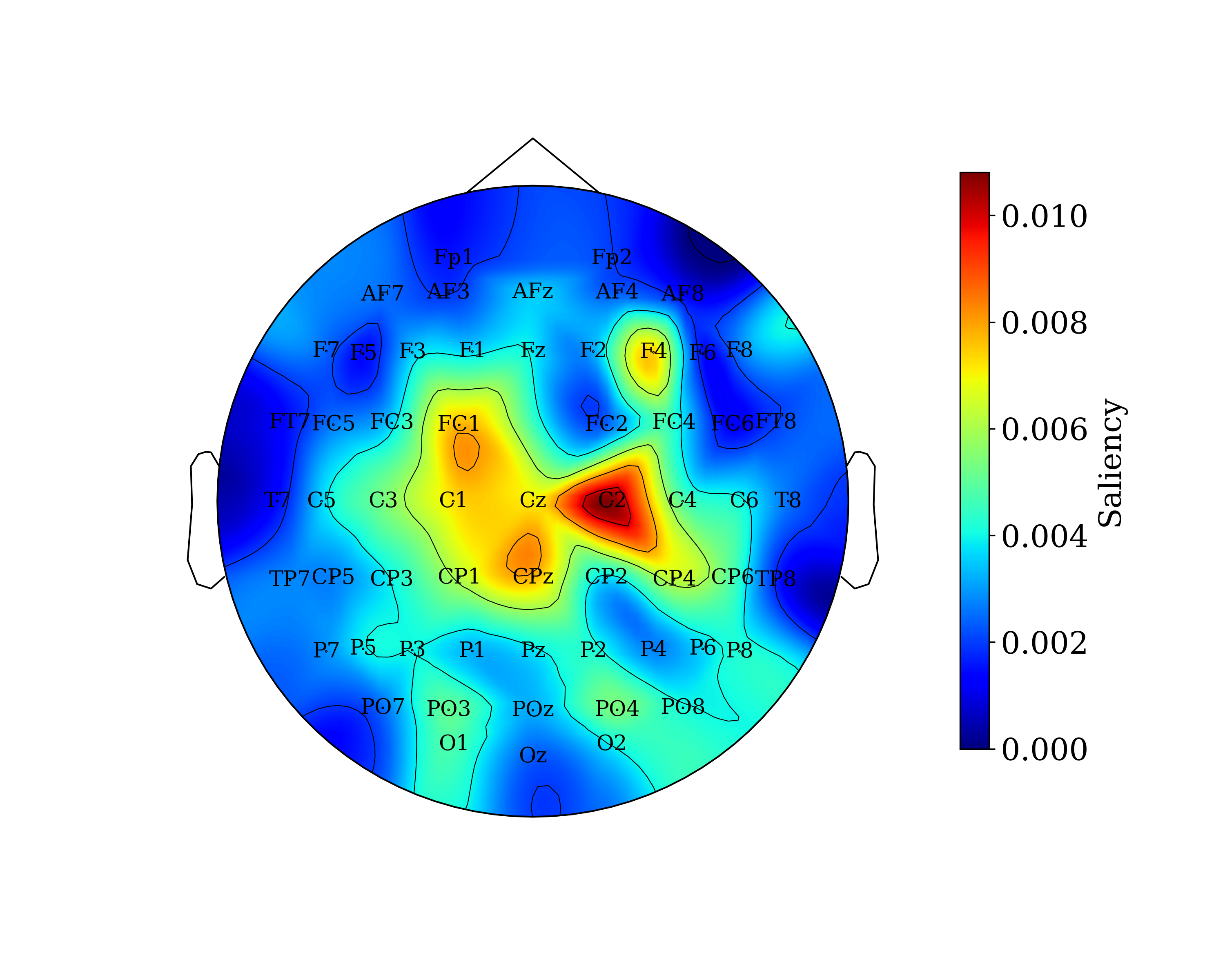}
    \caption{Saliency maps generated by NeuroDyGait computed across all folds.}

  \label{fig:saliency}
\end{figure}

The saliency maps from the test set are plotted in Fig.~\ref{fig:saliency}, showing the averaged channel-wise importance across all subjects during the walking task. The results indicate that the highlighted EEG channels were predominantly concentrated in the central sensorimotor areas. Notably, electrodes such as Cz, C1, C2, CPz, CP1, and FC1 exhibited the highest saliency, suggesting their critical role in gait-related neural processing. 

These observations reveal that the decoding model places strong emphasis on central regions typically associated with lower-limb motor control.

\subsection{Interpretability and Visualization}

To further elucidate the internal behavior of \modelname, we examined (1) the domain-head attention patterns learned by the fusion module and (2) the spatial saliency maps associated with gait-phase prediction.

\begin{figure}[htbp]
    \centering
    \includegraphics[width=\linewidth]{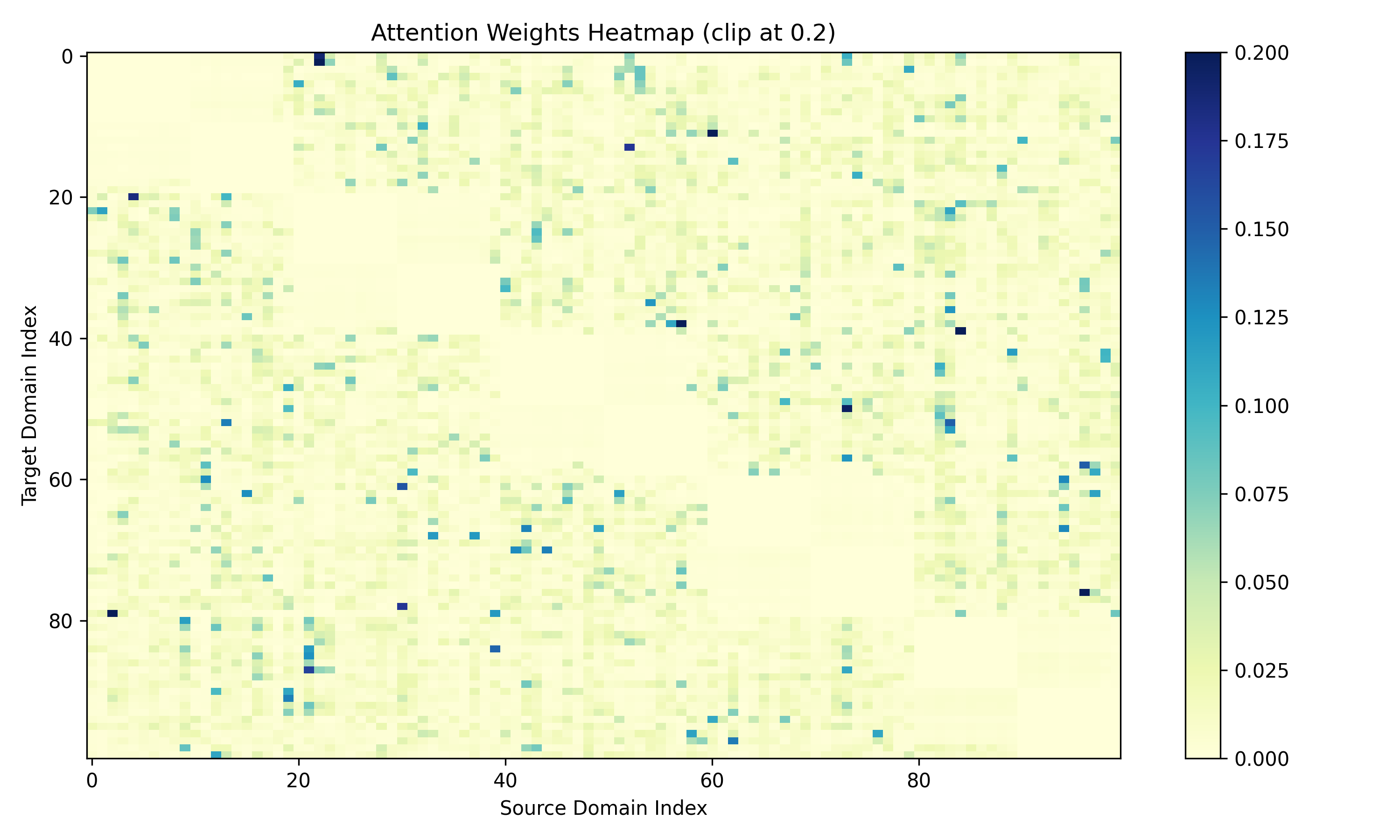}
    \caption{\textcolor{\differencecolor}{Aggregated attention map shows that each session learns a distinct weighting pattern, indicating that the model captures session-specific domain relationships rather than converging to a uniform structure.}}
    \label{fig:attn_vis}
\end{figure}

\textcolor{\differencecolor}{Figure~\ref{fig:attn_vis} shows the aggregated 100$\times$100 domain-attention matrix across all folds. A clear block-wise structure emerges: domains belonging to the same subject, or subjects sharing similar EEG statistics, consistently assign higher weights to each other. These subject-dependent patterns indicate that the model is not merely learning session identifiers but is capturing meaningful cross-domain neural similarity shaped by each subject’s characteristic EEG distribution. This behavior provides additional evidence that the multi-head domain fusion mechanism learns structured, subject-specific neural representations rather than relying on spurious correlations.}

\begin{figure}[htbp]
    \centering

    \begin{minipage}{0.45\linewidth}
        \centering
        \includegraphics[width=\linewidth]{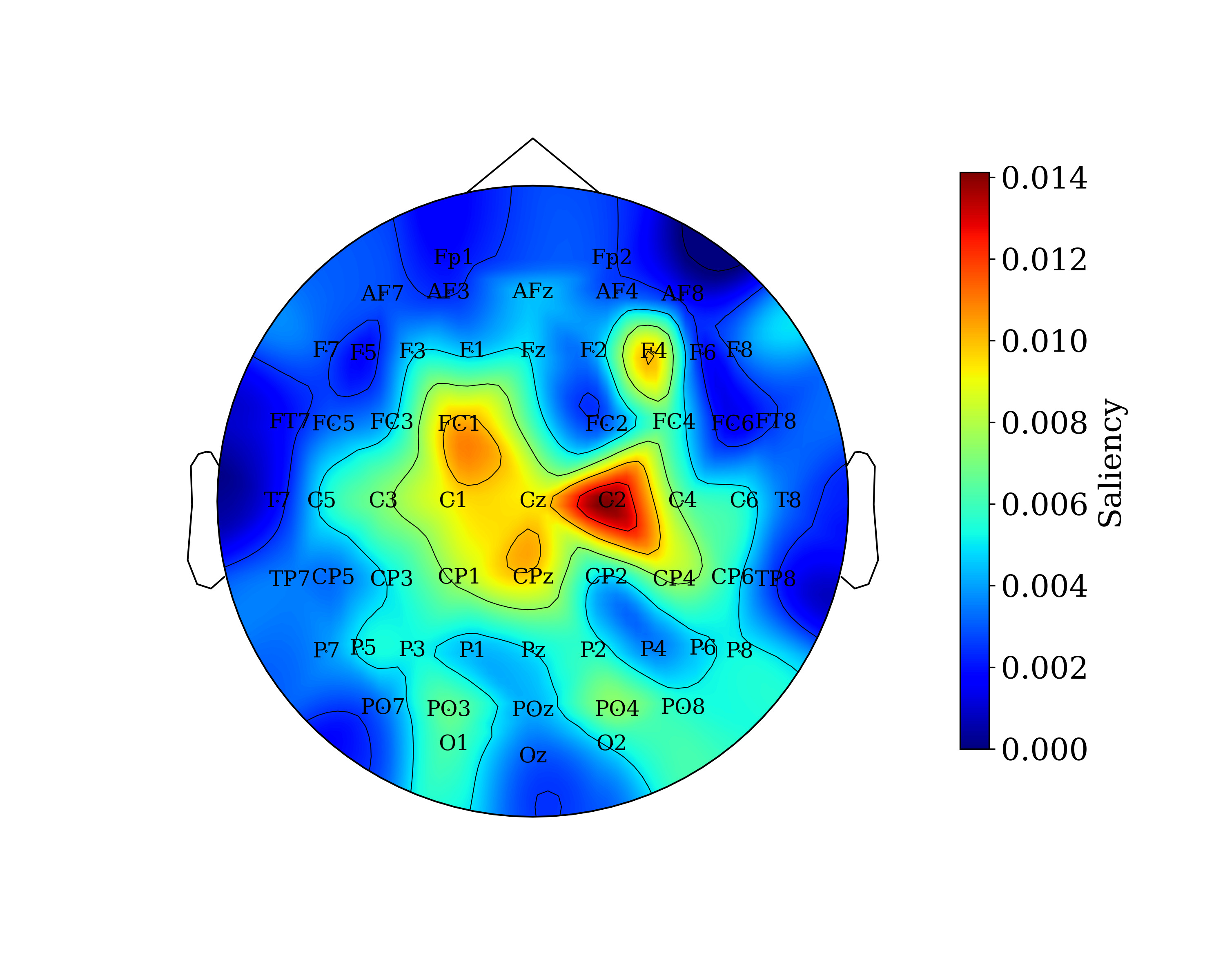}
        \vspace{0.1cm}
        \textbf{Phase 1}
    \end{minipage}
    \hfill
    \begin{minipage}{0.45\linewidth}
        \centering
        \includegraphics[width=\linewidth]{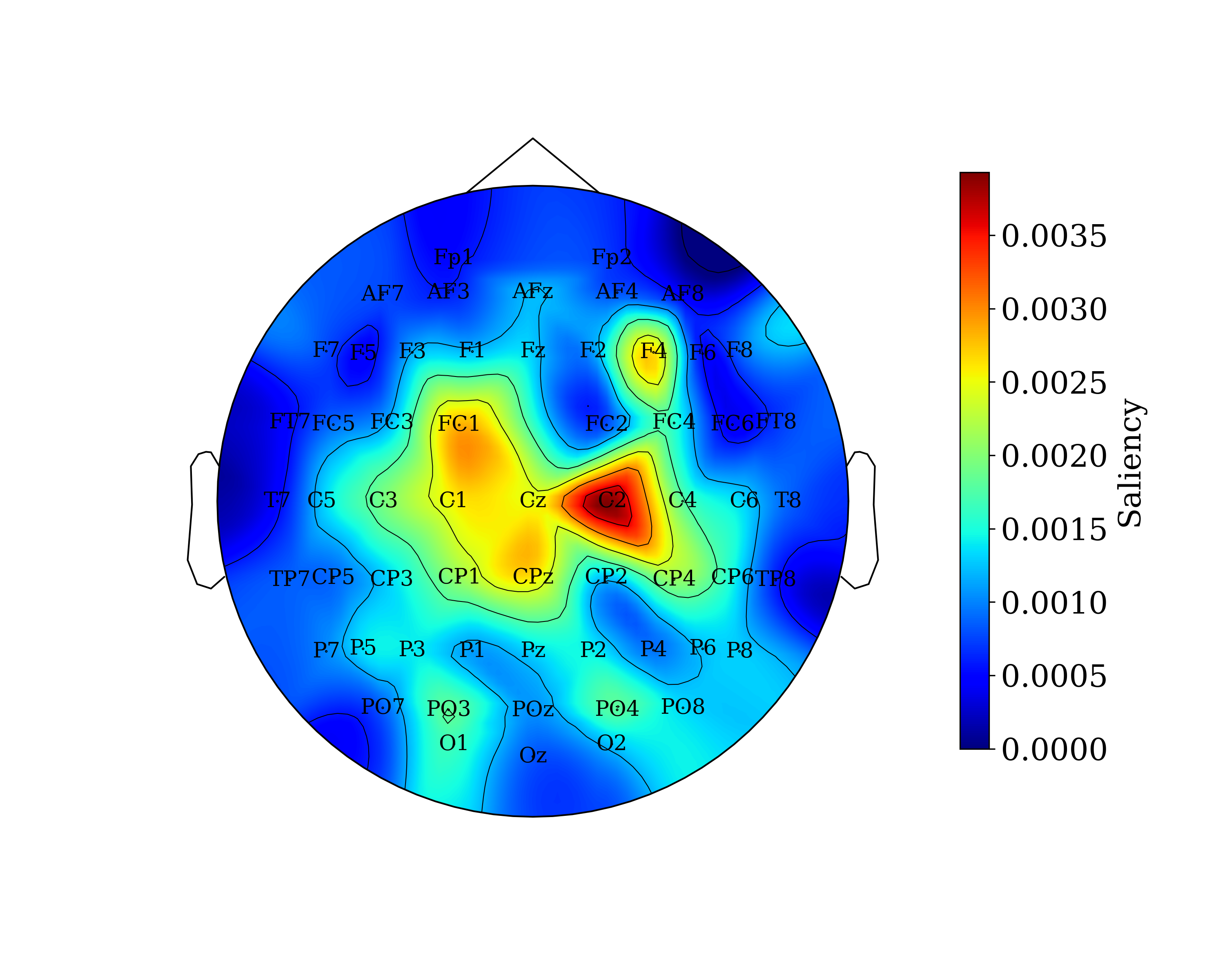}
        \vspace{0.1cm}
        \textbf{Phase 2}
    \end{minipage}

    \vspace{0.3cm}

    \begin{minipage}{0.45\linewidth}
        \centering
        \includegraphics[width=\linewidth]{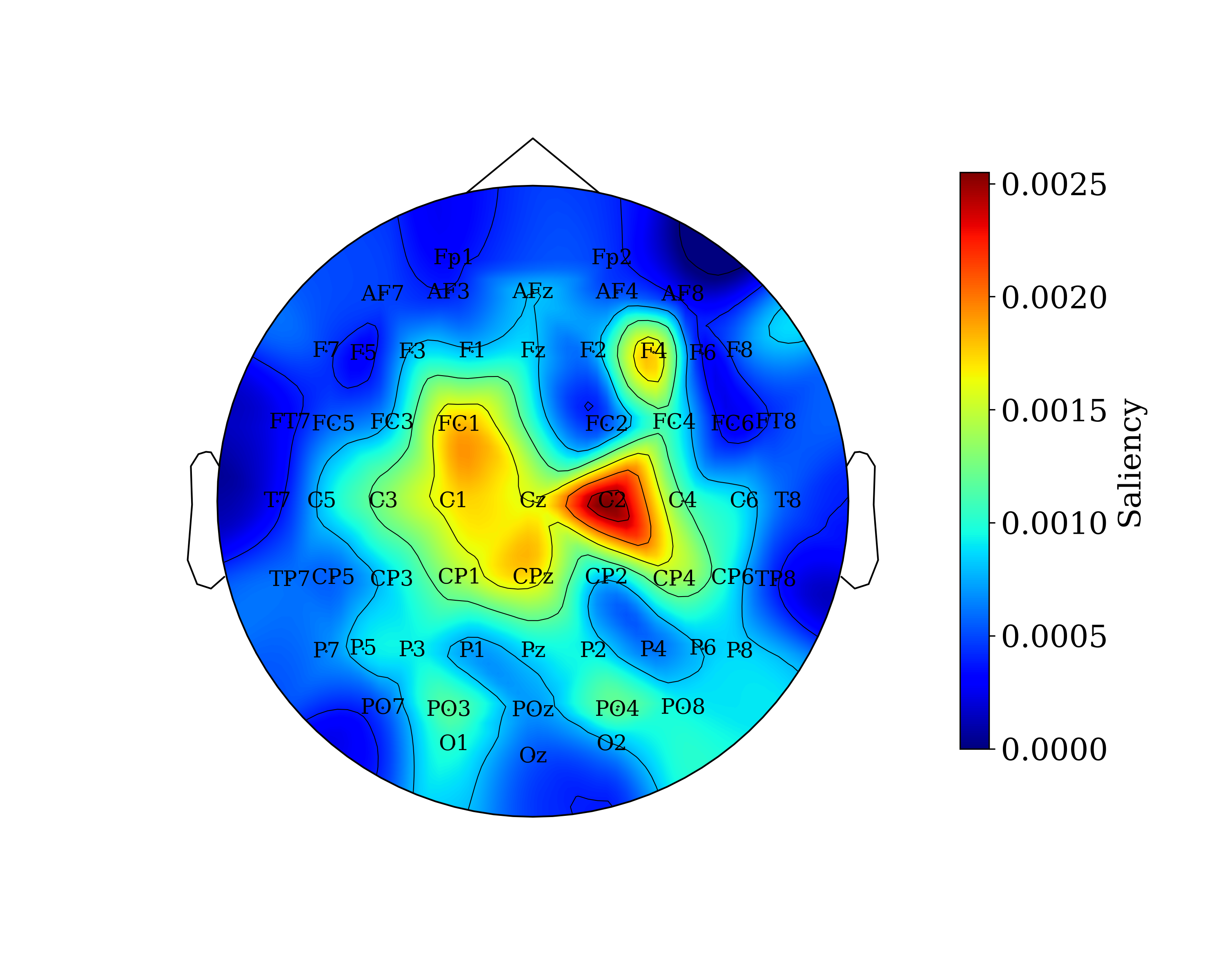}
        \vspace{0.1cm}
        \textbf{Phase 3}
    \end{minipage}
    \hfill
    \begin{minipage}{0.45\linewidth}
        \centering
        \includegraphics[width=\linewidth]{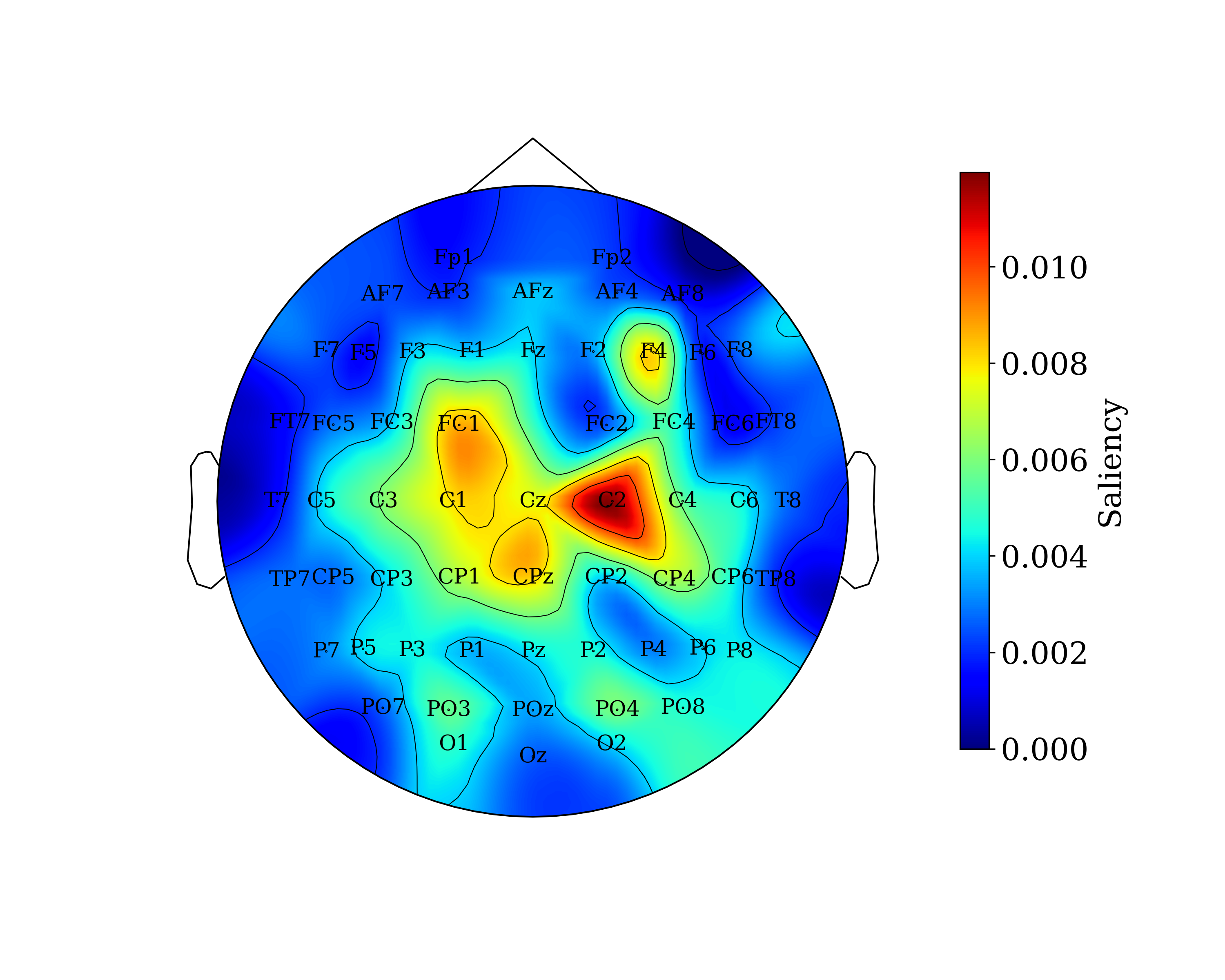}
        \vspace{0.1cm}
        \textbf{Phase 4}
    \end{minipage}

    \caption{\textcolor{\differencecolor}{Phase-specific cortical saliency maps. Spatial distributions are largely consistent across phases, reflecting stable neural activation during steady-state gait.}}
    \label{fig:phase_saliency}
\end{figure}

\textcolor{\differencecolor}{Figure~\ref{fig:phase_saliency} presents the saliency maps computed for four gait phases. Despite minor amplitude differences around central and centro-parietal electrodes (e.g., Cz, CP1, CP2), the overall spatial patterns remain highly consistent across phases. This stability is expected in steady-state walking, where the motor cortical rhythm supporting gait-cycle progression remains relatively unchanged across repeated cycles. Consequently, the phase saliency maps indicate that the encoder focuses on a stable set of motor-relevant sensors rather than phase-specific noise fluctuations.}

\subsection{Ablation Studies}

To understand the contribution of individual components in our framework, we conducted ablation experiments on GED and \textcolor{\differencecolor}{FMD} by selectively removing or modifying key loss terms and architectural modules. The results are summarized in Table~\ref{tab:ablation} and Table~\ref{tab:ablation_FMD}. These results collectively highlight the importance of each design component in our framework. In particular, the combination of contrastive pretraining with reconstruction and prediction, and dynamic domain fusion proves essential for learning transferable, motor-relevant EEG features and achieving strong decoding performance.

\begin{table}[htbp]
\centering
\caption{Ablation study results on GED dataset.}
\label{tab:ablation}
\scriptsize
\setlength{\tabcolsep}{4pt}
\resizebox{\linewidth}{!}{
\begin{tabular}{lccc}
\toprule
\textbf{Configuration} & \textbf{Pearson $r$} $\uparrow$ & \textbf{$R^2$ Score} $\uparrow$ & \textbf{RMSE} $\downarrow$ \\
\midrule
Full \modelname &  \textbf{{0.6980 $\pm$ 0.0742}} &\textbf{ {0.4847 $\pm$ 0.1085}} & \textbf{{0.7329 $\pm$ 0.0809} } \\

w/o Prediction Loss & 0.6862 $\pm$ 0.0758 & 0.4691 $\pm$ 0.1093 & 0.7475 $\pm$ 0.0839 \\

w/o Rel. Contrastive Loss & 0.6814 $\pm$ 0.0745 & 0.4617 $\pm$ 0.1069 & 0.7518 $\pm$ 0.0851 \\

w/o Reconstruction Loss & 0.6749 $\pm$ 0.0761 & 0.4549 $\pm$ 0.1066 & 0.7563 $\pm$ 0.0867 \\

w/o Stage I  & 0.6698 $\pm$ 0.0734 & 0.4513 $\pm$ 0.1098 & 0.7589 $\pm$ 0.0882 \\

w/o Multi-head Fusion & 0.6888 $\pm$ 0.0735 & 0.4725 $\pm$ 0.1076 & 0.7443 $\pm$ 0.0824 \\

\textcolor{\differencecolor}{Cross-attention $\rightarrow$ Cosine Similarity} 
    & \textcolor{\differencecolor}{0.6933 $\pm$ 0.0682} 
    & \textcolor{\differencecolor}{0.4811 $\pm$ 0.1023} 
    & \textcolor{\differencecolor}{0.7355 $\pm$ 0.0837} \\
\bottomrule
\end{tabular}
}
\end{table}

\begin{table}[htbp]
\centering
\caption{Ablation study results on \textcolor{\differencecolor}{FMD}.}
\label{tab:ablation_FMD}
\scriptsize
\setlength{\tabcolsep}{4pt}
\resizebox{\linewidth}{!}{
\begin{tabular}{lccc}
\toprule
\textbf{Configuration} & \textbf{Pearson $r$} $\uparrow$ & \textbf{$R^2$ Score} $\uparrow$ & \textbf{RMSE} $\downarrow$ \\
\midrule
Full \modelname & \textbf{0.2945 $\pm$ 0.1131} & \textbf{0.0485 $\pm$ 0.0342} & \textbf{0.9743 $\pm$ 0.0185} \\

w/o Prediction Loss & 0.2839 $\pm$ 0.1297 & 0.0375 $\pm$ 0.0419 & 0.9851 $\pm$ 0.0263 \\

w/o Rel. Contrastive Loss & 0.2734 $\pm$ 0.1314 & 0.0306 $\pm$ 0.0426 & 0.9942 $\pm$ 0.0271 \\

w/o Reconstruction Loss & 0.2689 $\pm$ 0.1289 & 0.0271 $\pm$ 0.0397 & 0.9980 $\pm$ 0.0284 \\

w/o Stage I  & 0.2651 $\pm$ 0.1253 & 0.0250 $\pm$ 0.0389 & 1.0023 $\pm$ 0.0297 \\

w/o Multi-head Fusion & 0.2791 $\pm$ 0.1308 & 0.0338 $\pm$ 0.0413 & 0.9893 $\pm$ 0.0259 \\

\textcolor{\differencecolor}{Cross-attention $\rightarrow$ Cosine Similarity} 
& \textcolor{\differencecolor}{0.2896 $\pm$ 0.1278} 
& \textcolor{\differencecolor}{0.0479 $\pm$ 0.0407} 
& \textcolor{\differencecolor}{0.9812 $\pm$ 0.0276} \\
\bottomrule
\end{tabular}
}
\end{table}

\textcolor{\differencecolor}{Furthermore, we evaluated a variant that replaces our cross-attention-based distance metric in Stage~I with a fixed cosine similarity. This ablation directly tests whether the proposed relation-aware distance contributes beyond a conventional similarity measure. Across both GED and \textcolor{\differencecolor}{FMD}, the cosine-similarity variant performs reasonably well but consistently lags behind the full model, demonstrating that the attention-based distance captures richer subject-specific and phase-relevant relationships. These observations confirm that the learned metric is a crucial component for robust cross-subject alignment and decoding accuracy.}

\subsection{Latency and Hardware Configuration}
\label{sec:latency}

\textcolor{\differencecolor}{To assess the real-time feasibility and deployment readiness of \modelname, we evaluated both the computational latency and the hardware resources required during training and inference.}

\paragraph{Training Hardware}
\textcolor{\differencecolor}{All Stage~I and Stage~II experiments were conducted on a workstation equipped with four NVIDIA A100 GPUs. Multi-GPU training was used solely to accelerate experimentation; however, the full pipeline can be trained on a single A100 GPU without modifications to the architecture or batch setup.}

\paragraph{Inference Latency}
\textcolor{\differencecolor}{Although each training sample uses a 2-second EEG window, the decoding process employs a sliding window with a 1.95-second overlap, resulting in an effective update interval of 50\,ms (20\,Hz). Thus, only the initial 2 seconds of EEG are required for warm-up, after which the model produces predictions continuously at 20\,Hz.}

\textcolor{\differencecolor}{Profiling on a single NVIDIA A100 GPU shows that the forward-pass latency per 2-second window is under 5\,ms, far below the 50\,ms sampling interval at 20\,Hz, confirming that \modelname\ comfortably satisfies real-time constraints for closed-loop BCI operation.}

\paragraph{Implications for Lightweight Deployment}
\textcolor{\differencecolor}{Given its low-latency forward pass and modest memory footprint, \modelname\ is suitable for deployment on lightweight or embedded GPU platforms. The model does not require multi-GPU resources during inference, and its real-time capability supports integration into gait-assistive and neurorehabilitation systems.}

\section{Discussion}

Our proposed framework, \modelname, demonstrates robust cross-subject and cross-dataset decoding of lower-limb kinematics from EEG. By leveraging domain-invariant neural patterns, \modelname\ adapts effectively to varied populations and recording conditions, a crucial property for practical deployment.

Pretraining results show that the encoder learns transferable motor representations that generalize across datasets. When pretrained on GED and transferred to \textcolor{\differencecolor}{FMD}, \modelname\ outperforms random initialization and accelerates convergence, highlighting the encoder’s ability to capture stable motor-relevant structure. \textcolor{\differencecolor}{The reduced zero-shot performance on FMD (GED→FMD $r{=}0.298$) is primarily attributable to two dataset-level factors: (i) the substantially smaller training cohort in FMD (10 subjects) and (ii) the broader diversity of locomotor conditions, including level, ramp, and stair walking. Downsampling GED to a comparable number of subjects yields a similar degradation, confirming that training-scale and task-diversity limitations constrain cross-domain generalization.}

At the same time, \textcolor{\differencecolor}{FMD imposes several subject-specific challenges that further explain why fine-tuning leads to large performance gains. First, FMD exhibits pronounced cross-subject domain shifts in EEG amplitude distributions, neuromuscular recruitment strategies, and gait coordination patterns, making the EEG–kinematic mapping highly individualized. Second, although the pretrained encoder provides stable phase-aware representations, the regression head must translate these embeddings into subject-specific joint trajectories; this mapping varies substantially across individuals and therefore benefits from light adaptation. Third, systematic timing misalignments—such as variations in step timing, neuromotor delay, or phase-transition boundaries—cause zero-shot predictions to have correct shapes but misaligned phase or amplitude. A small amount of fine-tuning effectively corrects these temporal and scaling offsets, explaining the dramatic improvement from $r{=}0.298$ to $r{=}0.617$. Together, these factors indicate that zero-shot limitations arise from subject heterogeneity rather than insufficient representational generality.}

The learned embeddings exhibit clear phase-specific organization, as shown by t-SNE visualizations. This indicates that the phase-aware contrastive objective effectively captures the temporal regularities of the gait cycle. \textcolor{\differencecolor}{Such structured embeddings provide clinically interpretable markers of gait-phase progression and may facilitate phase-synchronized neurorehabilitation interventions, such as real-time gait correction or exoskeleton control.} The consistency of phase structure across subjects further demonstrates that \modelname\ extracts causal temporal patterns rather than relying on superficial correlations.

The domain-relation mechanism enhances robustness by selectively weighting source sessions according to their relevance. The observed correlation between attention entropy and prediction accuracy suggests that confident and well-focused domain fusion contributes to improved generalization. \textcolor{\differencecolor}{This adaptability is particularly valuable when target distributions drift across subjects or sessions, enabling the model to emphasize physiologically compatible domains during inference.}

Spatial saliency analysis shows that \modelname\ consistently attends to motor-related cortical regions, including channels around Cz, C1, C2, CPz, CP1, and FC1, corresponding to the somatotopic representation of lower limbs. This alignment with established motor neurophysiology indicates that the model relies on meaningful cortical signals rather than spurious noise. \textcolor{\differencecolor}{Importantly, these salient patterns remain localized even during irregular gait segments, suggesting resilience to transient artifacts and supporting potential translational use in real-time BCI systems.}

\textcolor{\differencecolor}{Ablation experiments further validate the contributions of key components. Removing any Stage~I objective—reconstruction, prediction, or relative contrastive learning—reduces decoding accuracy, confirming their complementary roles in shaping invariant and semantically structured EEG embeddings.} Eliminating Stage~I altogether results in a substantial drop in performance, demonstrating that EEG-only supervision is insufficient for learning discriminative kinematic features. The multi-head domain-fusion module also proves essential for robust generalization. \textcolor{\differencecolor}{Consistent with the revised analysis, we observed that jointly optimizing the reconstruction, prediction, and relative contrastive losses with equal weights yields stable convergence without conflicting gradients and provides complementary improvements in representation quality.} \textcolor{\differencecolor}{We further found that varying the embedding dimension $d$ within a reasonable range (64–256) changed decoding performance by less than 2\%, indicating that \modelname\ is not sensitive to this hyperparameter.} Additionally, replacing the adaptive cross-attention distance with cosine similarity yields consistent but moderate degradation, indicating the benefit of learnable inter-domain re-weighting.

Although current evaluations focus on healthy individuals, the strong cross-domain robustness and efficient fine-tuning suggest that \modelname\ holds promise for clinical applications, including post-stroke or spinal cord injury rehabilitation. \textcolor{\differencecolor}{Future work should explore personalization strategies for impaired populations and investigate multimodal extensions—such as integrating EEG, EMG, and kinematics—to provide stable and informative representations under missing-modality or nonstationary conditions. We also envision evolving \modelname\ toward a multimodal physiological foundation model capable of continual adaptation across subjects, sessions, and motor tasks.}

\section{Conclusion}
In this work, we introduced \modelname, a dual-stage and domain-generalization framework for decoding lower-limb kinematics from EEG. By combining a learnable cross-modal distance metric with relative contrastive learning, and a domain attention module for fusion-based decoding, the model demonstrates robust generalization across subjects and datasets without requiring subject-specific calibration. Extensive evaluations show that \modelname\ effectively captures transferable and temporally structured motor representations, as evidenced by its performance and interpretability in both time and spatial domains.

The integration of saliency analysis and contrastive embeddings provides further insight into the model’s internal representations, revealing alignment with neurophysiological principles of motor control. The identification of central motor regions as key contributors to decoding reinforces the relevance of the learned features, offering a basis for potential clinical translation.

While the current validation is limited to healthy individuals, the demonstrated cross-domain robustness suggests strong potential for extension to neurorehabilitation settings. Future work should focus on evaluating the model in clinical populations, exploring multimodal extensions, and examining real-time deployment scenarios. Overall, \modelname\ contributes to the growing body of research at the intersection of brain–computer interfaces and rehabilitative neuroscience by offering a scalable and interpretable approach for neural decoding in ambulatory motor tasks.
\textcolor{\differencecolor}{In future extensions, we plan to incorporate multimodal physiological signals, subject-adaptive continual learning, and closed-loop robotic interfaces to enable fully autonomous rehabilitation feedback. Together, these efforts will transform NeuroDyGait into a foundation model for generalizable and interpretable brain–gait interaction.}

\bibliographystyle{elsarticle-num}
\bibliography{main}

\end{document}